\documentstyle[12pt,fullpage]{article}

\begin{document}
\parskip=10pt
\begin{flushright}
TAUP-2150-94
\end{flushright}
\vskip 3 true cm

\begin{center}
{\bf The Zeeman Effect for the Relativistic Bound State}

M. C. Land  and L. P. Horwitz

School of Physics and Astronomy  \\
Raymond and Beverly Sackler Faculty of Exact Sciences  \\
Tel Aviv University, Ramat Aviv, Israel

\end{center}
\vskip .5 true cm
\baselineskip 7mm
\parindent=0cm

\begin{abstract}
\parindent=0cm
\parskip=10pt
In the context of a relativistic quantum mechanics with
invariant evolution parameter, solutions for the relativistic
bound state problem have been found, which yield a spectrum for
the total mass
coinciding with the nonrelativistic Schr\"odinger energy spectrum.
These spectra were obtained by choosing an arbitrary spacelike
unit vector $n_\mu$ and restricting the support of the
eigenfunctions in spacetime to the subspace of the Minkowski
measure space, for which
$(x_\perp )^2 = [x-(x \cdot n) n ]^2 \geq 0$.  In this paper,
we examine the Zeeman effect for these bound states, which
requires $n_\mu$ to be a dynamical quantity.  We recover the
usual Zeeman splitting in a manifestly covariant form.

\end{abstract}
\section{Introduction}
\setcounter{equation}{0}

It has been shown \cite{I} that the replacement
\begin{equation}
r = \sqrt{({\bf r}_1 - {\bf r}_2)^2} \qquad \longrightarrow \qquad
\rho = \sqrt{({\bf r}_1 - {\bf r}_2)^2 - (t_1 -t_2)^2}
\label{eqn:1.1}
\end{equation}
in the argument of the usual central force potentials of
non-relativistic mechanics leads to a relativistic problem,
yielding a mass spectrum coinciding with the nonrelativistic
Schr\"odinger energy spectrum, in the context of a relativistic
quantum mechanics with invariant parameter \cite{rqm}
(the correspondence is established by the fact that
$t_1 \rightarrow t_2 $ in the
nonrelativistic limit).  These spectra are obtained when one
chooses a spacelike unit vector $n_\mu$ ($g_{\mu\nu} = {\rm
diag}(-1,1,1,1)\; \Rightarrow \; n^2=+1$) and restricts the
support of the eigenfunctions in spacetime to the subspace of
the Minkowski measure space corresponding to the condition
\begin{equation}
(x_\perp )^2 = [x-(x \cdot n) n ]^2 \geq 0,
\label{eqn:1.2}
\end{equation}
where we denote by $x \equiv x^{\mu}$ the {\it relative} coordinates
$x^\mu_1 - x^\mu_2$,
for the two body system, and $x^2 = x^\mu x_\mu$.  The restricted
space, called the RMS
(Restricted Minkowski Space), is transitive and invariant under
the O(2,1) subgroup of O(3,1) leaving $n_\mu$ invariant
and translations along $n_\mu$.

The two-body (Poincar\'e invariant) Hamiltonian in this theory,
\begin{equation}
 K = {p_{1\mu}p_1^{\mu} \over {2M_1}} + {p_{2 \mu} p_2^{\mu}
\over {2M_2}} + V(\rho),
\label{eqn:1.3}
\end{equation}
is quadratic in the
four momenta, and one may separate variables of the center of mass
motion and relative motion in the same way as in the
nonrelativistic theory,
\begin{equation}
K = {P^{\mu} P_{\mu} \over 2M} + {p^{\mu} p_{\mu}
\over 2m} + V(\rho),
\label{eqn:1.4}
\end{equation}
where
\begin{equation}
P^\mu = p_1^\mu + p_2^\mu \qquad\qquad M = M_1 + M_2
\label{eqn:1.5}
\end{equation}
$$p^\mu = (M_2p_1^\mu - M_1 p_2^\mu)/M \qquad\qquad m=M_1M_2/M.$$
In \cite{I}, $n_\mu$ was chosen to be the $z$-axis, and the relative
Hamiltonian
\begin{equation}
K_{rel} = {p^{\mu} p_{\mu} \over 2m} + V(\rho)
\label{eqn:1.6}
\end{equation}
was expressed in terms of coordinates with the parameterization
$$y^0 = \rho \; \sinh \beta \; \sin \theta
\qquad\qquad y^1 = \rho \; \cosh \beta \; \sin \theta \; \cos \phi$$
\begin{equation}
y^2 = \rho \; \cosh \beta \; \sin \theta \; \sin \phi
\qquad\qquad y^3 = \rho \; \cos \theta
\label{eqn:1.7}
\end{equation}
for which
\begin{equation}
(y^1)^2 + (y^2)^2 - (y^0)^2 \geq 0.
\label{eqn:1.8}
\end{equation}

It was shown in \cite{I,II} that the eigenfunctions of $K_{rel}$
form irreducible representations of SU(1,1) --- in the double
covering of O(2,1) --- parameterized by the spacelike vector
$n_\mu$ stabilized by the particular O(2,1).  In \cite{II}, an
induced representation of SL(2,C) was constructed, by applying
the Lorentz group to the RMS coordinates $x^\mu$ and the frame
orientation $n_\mu$, and studying the action on these
wavefunctions.  One first observes that wavefunctions with
support on
\begin{equation}
x \in {\rm RMS}(n_\mu) = \left\{ x \ | \ [x-(x \cdot n) n
]^2 \geq 0 \right\}
\label{eqn:1.9}
\end{equation}
may be written as functions
of $n_\mu$ and the coordinates of a standard frame $y \in {\rm
RMS}({\mathaccent'27n}_\mu)$ since, given the Lorentz transformation ${\cal L}$
such that $ {\mathaccent'27n} = {\cal L} (n) \; n $, it follows that
\begin{equation}
x \in {\rm RMS}(n_\mu) \qquad {\rm and} \qquad y = {\cal L} (n) \;
x \qquad \Longrightarrow \qquad y \in {\rm RMS}({\mathaccent'27n}_\mu).
\label{eqn:1.10}
\end{equation}
By choosing ${\mathaccent'27n} = (0,0,0,1)$ as in \cite{I}, the
parameterization (\ref{eqn:1.7}) may be used for $y^\mu$.  Now,
under Lorentz transformations labeled by $\Lambda$, the wavefunctions
were shown to transform as
\begin{equation}
\psi_n (y) \rightarrow \psi_n^\Lambda (y) = \psi_{\Lambda^{-1}n}
(D^{-1} (\Lambda , n) \; y)
\label{eqn:1.11}
\end{equation}
where $\Lambda$ acts directly on $n_\mu$.  The representations
are moved on an orbit generated by this spacelike vector,
and the Lorentz transformations act on $y^\mu$ through the O(2,1)
little group, represented by $D^{-1} (\Lambda , n)$, with the
property
\begin{equation}
D^{-1} (\Lambda , n) \; {\mathaccent'27n} = {\cal L} (\Lambda n) \;
\Lambda \; {\cal L}^T (n) \; {\mathaccent'27n} \equiv {\mathaccent'27n} .
\label{eqn:1.12}
\end{equation}
The matrix ${\cal L} ^T(n)$ was chosen in \cite{II} to be a boost in
the three-direction, a rotation about the two-axis, followed by
a rotation about the one-axis.  Thus,
\begin{equation}
{\cal L} ^T(n)  = e^{\gamma {\cal M}^{23}}e^{\omega
{\cal M}^{31}}e^{\alpha {\cal M}^{03}}
\label{eqn:1.13}
\end{equation}
where
\begin{equation}
({\cal M}^{\sigma\lambda})^{\mu\nu} = g^{\sigma\mu}g^{\lambda\nu} -
g^{\sigma\nu}g^{\lambda\mu},
\label{eqn:1.14}
\end{equation}
and so
\begin{equation}
{\cal L} ^T(n)  = \pmatrix{\cosh \alpha & 0 & 0 & \sinh \alpha \cr
- \sin\omega\; \sinh\alpha & \cos\omega & 0 & - \sin\omega\; \cosh\alpha
\cr \sin\gamma \; \cos\omega \; \sinh\alpha & \sin\gamma \; \sin\omega
& \cos\gamma & \sin\gamma \; \cos\omega \; \cosh\alpha \cr \cos\gamma \;
\cos\omega \; \sinh\alpha & \cos\gamma \; \sin\omega & -\sin\gamma &
\cos\gamma \; \cos\omega \; \cosh\alpha \cr } ,
\label{eqn:1.15}
\end{equation}
which provides the parameterization of $n_\mu$ as
\begin{equation}
n_\mu = \pmatrix{\sinh \alpha \cr -\sin\omega\; \cosh\alpha \cr
\sin\gamma \; \cos\omega \; \cosh\alpha \cr \cos\gamma \; \cos\omega \;
\cosh\alpha \cr } .
\label{eqn:1.16}
\end{equation}
By examining the generators $h_{\alpha\beta}(n)$ of
(\ref{eqn:1.11}), which form a representation of the O(3,1) Lie
algebra (through their action on $y$ and $n$), the Casimir
operators
\begin{equation}
\hat{c}_1= {1\over 2} h_{\alpha\beta}(n) h^{\alpha\beta}(n) \qquad \hat{c}_2 =
{1\over 2} \epsilon^{\alpha\beta\gamma\delta} h_{\alpha\beta}(n)
h_{\gamma\delta}(n)
\label{eqn:1.17}
\end{equation}
as well as the operators of the SU(2) subgroup
\begin{equation}
{\bf L}^2 (n) = {1\over 2} h_{ij}(n)h^{ij}(n) \qquad L_1 (n)
= h^{23} (n) = -i {\partial \over \partial \gamma}
\label{eqn:1.18}
\end{equation}
can be constructed as a commuting set.  Moreover, the operator
\begin{equation}
\Lambda = {1 \over 2}M^{\mu \nu} M_{\mu \nu} \rightarrow
\ell(\ell +1) - {3\over 4},
\label{eqn:1.19}
\end{equation}
where $M^{\mu \nu} = y^\mu p^\nu - y^\nu p^\mu$, and the O(2,1)
Casimir $N^2 = (M^{01})^2 + (M^{02})^2 + (M^{12})^2 $
commute with this set and, wavefunctions were
constructed which are eigenfunctions of the set
\begin{equation}
\{\Lambda , N^2, \hat{c}_1, \hat{c}_2 , {\bf L}^2 (n) , L_1 (n) \}
\label{eqn:1.20}
\end{equation}
with eigenvalues $Q= \{ \ell(\ell +1) - {3 \over 4}, n^2
-{1\over 4 }, c_1, c_2, L(L+1), q \}$.  The requirement that
these wavefunctions lie in a unitary irreducible representation of
SL(2,C) (they are in the principal series), imposes the
condition $c_1 = \hat{n} ^2 -1 - c_2^2 / \hat{n} ^2 $, where $\hat{n} =
n+1/2$.

The remaining ``radial'' function, after the transformation
$\hat R (\rho) = R (\rho) /\sqrt{\rho} $ of the radial part of
$\psi_n(y)$, then must satisfy an equation which is precisely of
the form of the nonrelativistic Schr\"odinger radial equation in three
dimensions (and has the same normalization).
The states $\psi_n (y)$ are then eigenstates of the Lorentz
invariant $K_{rel}$, whose support is on the RMS($n$), with the
quantum numbers (\ref{eqn:1.20}), and a principal quantum number
$n_a$.  In
particular, the solutions for the problem corresponding to the
Coulomb potential \cite{I}, yield bound states with a mass spectrum
which coincides with the nonrelativistic Schr\"odinger energy
spectrum.  The observed {\it energies} for such systems are
determined by the
values of $P^\mu P_\mu$, i.e., $-E^2$ in the center of momentum
frame; from (\ref{eqn:1.4}) one obtains, in an expansion in
orders of $1/c^2$, the nonrelativistic spectrum with
relativistic corrections.

The selection rules for dipole
radiation from these states have been calculated \cite{selrul}
and have been shown to be identical with those of the usual
nonrelativistic theory, expressed in a manifestly covariant
form,
\begin{equation}
\{ \Delta \ell = \pm 1; \;\; \Delta q =0, \pm 1\}.
\label{eqn:1.21}
\end{equation}
In addition to the transverse and longitudinal polarizations of
the nonrelativistic theory, there is a ``scalar'' transition,
induced by the relative time coordinate.  The ``scalar''
polarization and the longitudinal polarization induce the same
$\Delta q =0$ transition for
the relativistic case, which has a natural interpretation in terms of
the Gupta-Bleuler quantization of the photon.  This
relationship shows that the wavefunctions act correctly as
representations of the the Lorentz group.  Moreover, it was shown in
\cite{selrul} that the change in $q$, the eigenvalue of $L_1
(n)$, corresponds to a change in the orientation of $n_\mu$ with
respect to the polarization of the emitted or absorbed photon.
That the magnetic quantum number $q$ depends on the frame
orientation should not be surprising, because the operator $L_1
(n)$ belongs to the SU(2) subgroup of SL(2,C), and acts on
$n_\mu$, but not on the RMS coordinates (it was shown in
\cite{selrul} that for $\Lambda$ a rotation about the 1-axis,
${D^{-1} (\Lambda , n)} \equiv 1$).

In this paper, we provide a derivation of the Zeeman effect for
the bound states, which requires allowing $n_\mu$ to be come a
dynamical quantity.  We begin with a discussion of the
classical O(3,1) in the induced representation and obtain the
group generators, which coincide with those of \cite{II}, when the
momenta are understood as derivatives in the Poisson bracket
sense.  We
construct a classical Lagrangian, in which $n_\mu$ plays an
explicit dynamical role, and show that the generators are
conserved.  We then construct the Hamiltonian, which may be
unambiguously quantized and made locally gauge invariant.
Finally, it is shown that an external gauge field representing
a constant magnetic field induces an energy level splitting
corresponding to the usual nonrelativistic expression.

\section{The Configuration Space}
\setcounter{equation}{0}

We shall be interested, in this section, in the classical
relativistic mechanics of
events of spacelike separation.  We characterize the
separation vectors by the coordinates $(n,y)$, where $n$ is the
spacelike unit vector parameterized in (\ref{eqn:1.16}); $y \in
{\rm RMS}({\mathaccent'27n})$ is parameterized in (\ref{eqn:1.7}) (note that
${\cal L}^T (n) y \in {\rm RMS}(n)$) and satisfies (\ref{eqn:1.2}).

Under a Lorentz transformation $\Lambda$, we know that
\begin{equation}
n \rightarrow n' = \Lambda \; n \qquad x \rightarrow x' =
\Lambda \; x
\label{eqn:2.1}
\end{equation}
It follows from (\ref{eqn:1.10}) and (\ref{eqn:2.1}) that
\begin{equation}
x' = \Lambda \; x = \Lambda {\cal L} (n)^T \; y =
{\cal L} (\Lambda n)^T {\cal L} (\Lambda n)\; \Lambda \;
{\cal L} (n)^T \; y = {\cal L} (n')^T \; y' .
\label{eqn:2.2}
\end{equation}
Thus $y$ transforms as
\begin{equation}
y \rightarrow y' = {D^{-1} (\Lambda , n)} \; y ,
\label{eqn:2.3}
\end{equation}
where (as in (\ref{eqn:1.12})) ${D^{-1} (\Lambda , n)}={\cal L}
(\Lambda n)\; \Lambda \;{\cal L}(n)^{T} $ belongs to the O(2,1)
which leaves ${\mathaccent'27n}$ invariant,
i.e.,
\begin{equation}
D^{-1} (\Lambda , n) \; {\mathaccent'27n} = {\cal L} (\Lambda n)\;
\Lambda \; {\cal L} (n)^T \; {\mathaccent'27n} = {\mathaccent'27n} ,
\label{eqn:2.4}
\end{equation}
and hence the relation (\ref{eqn:1.8}) is preserved.
The coordinates thus transform as
\begin{equation}
\Lambda :  \; (n,y) \quad \rightarrow \quad (n,y)' = (\Lambda  n ,
D^{-1} (\Lambda , n) y).
\label{eqn:2.5}
\end{equation}

We wish now to construct a model for the Zeeman effect in
this covariant framework.  To do this, we recall that in the
computation of the selection rules for radiative processes,
as we remarked above, the restriction $\Delta q =0, \pm 1$
refers to a reaction of the radiation on the orientation of
the coset label $n^\mu$ of the induced representation.  In the
dipole approximation, the transition operator is $x^{\mu}$, and
in \cite{selrul}, we demonstrated that the conservation of the
eigenvalues $L$ and
$n$ in the matrix elements of $x^{\mu}$ implies the vanishing
of the matrix element $<\!  {\ell}'n'|\sin\theta|{\ell}n\!  >$,
leaving only the terms containing $<\! {\ell}'n|\cos\theta|{\ell}n\! >$
in the calculations.  Since this term arises only from the
$y^3 = \rho \; \cos \theta $ component of
$y^{\mu}$, the terms of $x^{\mu}$ which contribute to
these matrix elements are of the form ${\cal L}(n)^{3\mu}y_3$.
The 3-column of ${\cal L}^T$ is precisely $n_{\mu}$, so the calculation
factors as
\begin{eqnarray}
<n_{a'} {\ell}' n' L' q' c'_2 | x^{\mu} |n_{a}
{\ell} n L q c_2 > &=&  <n_{a'} {\ell}' n' L' q' c'_2 | \rho \;
\cos \theta \; n^{\mu} |n_{a} {\ell} n L q c_2 >
\nonumber \\
&=& <n_{a'}{\ell}' | \rho | n_a {\ell} >
 <{\ell}'n|\cos\theta|{\ell}n> <n' L' q' c'_2 | n^{\mu} |n L q c_2 >
\nonumber \\
&=&<n_{a'} {\ell}' | \rho | n_a {\ell} >
<{\ell}'n|\cos\theta|{\ell}n>
\; <n L q' c_2 | n^{\mu} |n L q c_2 > \nonumber \\
&&\mbox{\qquad} \times \delta_{nn'} \; \delta_{LL'} \;
\delta(c_2 - c'_2).
\label{eqn:2.6}
\end{eqnarray}
Since $| n_a {\ell} >$ refers to the radial functions and the
functions $|{\ell}n \!>$ are the usual spherical harmonics,
(\ref{eqn:2.6}) shows directly that it is the orientation of
$n_\mu$ which determines the transition in $q$.

We deduce from this result that the vector $n^\mu$ must be
effectively coupled to the radiation field, and we shall
build our model for coupling to the electromagnetic field by
adding to the Lagrangian a kinetic term for the evolution of
$n^\mu$ which, with minimal gauge invariance, provides the
Zeeman coupling.

The {\it velocity} $\dot n = dn/d\tau$ transforms just as $n$ does,
since $\tau$ is invariant:
\begin{equation}
n' =\Lambda \; n \quad \Longrightarrow \quad \dot n' =\Lambda \; \dot n
\label{eqn:2.7}
\end{equation}
but since ${\cal L} (n)$ is now $\tau$-dependent, the transformation of
$\dot y$ is more complicated.  We may write
\begin{eqnarray}
y= {\cal L} (n(\tau)) \; x \quad &\Longrightarrow& \quad \dot y =
{\cal L} (n) \dot x + \dot {\cal L} (n) x \label{eqn:2.8}\\
x= {\cal L} (n(\tau))^T \; y \quad &\Longrightarrow& \quad
\dot x = {\cal L} (n)^T \dot y + \dot {\cal L} (n)^T y
\label{eqn:2.9}
\end{eqnarray}
and we see that since $d/d\tau$ and the Lorentz
transformation commute, (\ref{eqn:2.8}) is, in fact, form
invariant:
\begin{eqnarray}
(\dot y)' &=& {\cal L} (n') \dot x' + \dot {\cal L} (n') x' \nonumber \\
&=& {\cal L} (\Lambda n) [\Lambda \dot x ] + \dot {\cal L} (\Lambda n)
[\Lambda x]
\nonumber \\
&=& {\cal L} (\Lambda \; n) \Lambda [{\cal L} (n)^T \dot
y + \dot {\cal L} (n)^T y ] + \dot {\cal L} (\Lambda n) [\Lambda {\cal
L}(n))^T
\; y ] \nonumber \\
&=& [ {\cal L} (\Lambda n) \Lambda {\cal L} (n)^T ] \dot y +
[ {\cal L} (\Lambda n) \Lambda \dot {\cal L} (n)^T  + \dot {\cal L}
(\Lambda \; n) \Lambda {\cal L} (n))^T] \; y
\nonumber \\
&=& D^{-1} (\Lambda , n) \dot y + \dot D^{-1} (\Lambda , n) \; y
\nonumber \\
&=& \frac{d}{d\tau} [ D^{-1} (\Lambda , n) \; y ] .
\label{eqn:2.10}
\end{eqnarray}
The {\it phase space} (which must include $n,\dot n$) transforms as:
\begin{equation}
\Lambda: \quad \{ (n,y);(\dot n,\dot y)\} \longrightarrow
\{ (\Lambda n, D^{-1} (\Lambda , n) y ); (\Lambda \dot n,
D^{-1} (\Lambda , n) \dot y+ \dot D^{-1} (\Lambda , n) y) \}.
\label{eqn:2.11}
\end{equation}

We now examine the generators of the Lorentz transformation
represented in (\ref{eqn:2.5}).  We take
\begin{equation}
\Lambda =1+ \lambda +o(\lambda ^2)
\label{eqn:2.12}
\end{equation}
and write $\lambda$ as
\begin{equation}
\lambda = \frac{1}{2} \; \omega_{\alpha\beta} \; {\cal M}^{\alpha\beta}
\label{eqn:2.13}
\end{equation}
where $\omega_{\alpha\beta}, \; \alpha,\beta =0, \cdots,3$
is (infinitesimal) antisymmetric.  The matrix generators
\begin{equation}
{\cal M}^{\alpha\beta} =  \left.  \frac{\partial \lambda}{\partial
\omega_{\alpha\beta}} \right|_{\omega=0}
\label{eqn:2.14}
\end{equation}
are those given in (\ref{eqn:1.14}).
According to (\ref{eqn:2.12}) and (\ref{eqn:2.13}), (\ref{eqn:2.5})
becomes
\begin{equation}
\Lambda :  \; (n,y) \quad \rightarrow \quad (n,y)' = (n+ \lambda
n , {\cal L}(n+\lambda n )(1+\lambda ) {\cal L}(n)^T y) + o(\omega ^2).
\label{eqn:2.15}
\end{equation}
Defining the generators of $\xi=(n,y) \rightarrow \xi'=(n',y')$ as
\begin{equation}
X_{\alpha\beta} = \sum_{i=1}^{8} \left. \frac{\partial
\xi^i}{\partial \omega^{\alpha\beta}} \right|_{\omega=0}
\frac{\partial}{\partial \xi^i}
\label{eqn:2.16}
\end{equation}
where for $i=1,\cdots,4$, $\xi^i = n^\mu$, $\mu=0,\cdots,3$,
and for $i=5,\cdots,8$, $\xi^i = y^\mu$, $\mu=0,\cdots,3$.
Thus, for $i=1,\cdots,4$,
\begin{eqnarray}
\left. \frac{\partial \xi^i}{\partial \omega^{\alpha\beta}}
\right|_{\omega=0} &=& \left. \frac{\partial }{\partial
\omega^{\alpha\beta}} (n^i + (\lambda n)^i ) \right|_{\omega=0}
\nonumber \\
&=& \left. \frac{\partial }{\partial \omega^{\alpha\beta}} (n^i +
(\frac{1}{2}\omega^{\sigma\rho}{\cal M}_{\sigma\rho} n)^i )
\right|_{\omega=0} \nonumber \\
&=&\frac{1}{2}(\delta^{\sigma}_{\alpha} \delta^{\rho}_{\beta} -
\delta^{\sigma}_{\beta} \delta^{\rho}_{\alpha} ) ({\cal M}_{\sigma\rho}
n)^i \nonumber \\
&=& ({\cal M}_{\alpha\beta})^i_{\ j} n^i ,
\label{eqn:2.17}
\end{eqnarray}
so that
\begin{eqnarray}
\sum_{i=1}^{4} \left. \frac{\partial \xi^i}{\partial
\omega^{\alpha\beta}} \right|_{\omega=0} \frac{\partial}{\partial \xi^i}
&=& ({\cal M}_{\alpha\beta})^\mu_{\ \nu} n^\nu \frac{\partial}{\partial n^\mu}
\nonumber \\
&=& (g^{\mu}_{\alpha} g_{\beta\nu} - g^{\mu}_{\beta} g_{\alpha\nu})
n^\nu \frac{\partial}{\partial n^\mu} \nonumber \\
&=&
n_\beta \frac{\partial}{\partial n^\alpha} -
n_\alpha \frac{\partial}{\partial n^\beta}
\label{eqn:2.18}
\end{eqnarray}
which was called $d(\lambda_{\alpha\beta})$ in \cite{II}.

Now for $i=5,\cdots,8$,
\begin{eqnarray}
\left. \frac{\partial \xi^i}{\partial \omega^{\alpha\beta}}
\right|_{\omega=0} &=& \left. \frac{\partial}{\partial
\omega^{\alpha\beta}} \left[{\cal L} (n+\lambda n) (1+\lambda)
{\cal L} (n)^T y \right] ^i \right|_{\omega=0} \nonumber \\
&=& \left[ \left. \frac{\partial}{\partial \omega^{\alpha\beta}}
{\cal L} (n+\lambda n) \right|_{\omega=0} {\cal L} (n)^T y \right] ^i
+ \left[ {\cal L} (n)
\left. \frac{\partial}{\partial \omega^{\alpha\beta}}
\lambda
\right|_{\omega=0} {\cal L} (n)^T y \right] ^i  \nonumber \\
&=& \left[ \frac{\partial}{\partial n^\mu} {\cal L} (n)  \left.
\frac{\partial}{\partial
\omega^{\alpha\beta}} (\lambda n)^\mu \right|_{\omega=0}
{\cal L} (n)^T y \right] ^i  + \left[ {\cal L} (n) {\cal M}_{\alpha\beta} {\cal
L}(n)^T y
\right] ^i
\nonumber \\
&=& \left[ - ({\cal M}_{\alpha\beta})^\mu_\nu n^\nu {\cal L} (n)
\frac{\partial}{\partial n^\mu} {\cal L} (n)^T  + {\cal L} (n)
{\cal M}_{\alpha\beta} {\cal L} (n)^T \right]^{ij} y_j
\label{eqn:2.19}
\end{eqnarray}
where we have used the fact that
\begin{equation}
{\cal L} (n) {\cal L} (n)^T =1 \quad \Longrightarrow \quad
\left(\frac{\partial}
{\partial n^\mu} {\cal L} (n) \right) {\cal L} (n)^T  + {\cal L} (n)
\frac{\partial}{\partial n^\mu} {\cal L} (n) ^T =0 .
\label{eqn:2.20}
\end{equation}
Thus, we find that
\begin{equation}
\sum_{i=5}^{8} \left. \frac{\partial \xi^i}{\partial
\omega^{\alpha\beta}} \right|_{\omega=0} \frac{\partial}{\partial
\xi^i} =  \left[ {\cal L} (n) {\cal M}_{\alpha\beta}
{\cal L} (n)^T - ({\cal M}_{\alpha\beta})^\mu_{\ \nu} n^\nu
{\cal L} (n) \frac{\partial}{\partial n^\mu} {\cal L} (n)^T
\right]^{\rho\sigma} y_\sigma \frac{\partial}{\partial y^\rho}
\label{eqn:2.21}
\end{equation}
Using (\ref{eqn:1.14}) for ${\cal M}_{\alpha\beta}$, we obtain
\begin{equation}
\sum_{i=5}^{8} \left. \frac{\partial \xi^i}{\partial
\omega^{\alpha\beta}} \right|_{\omega=0} \frac{\partial}{\partial
\xi^i} =  {\cal L} _{\sigma\beta} {\cal L} ^{\rho}_{\ \alpha} (y^\sigma
\frac{\partial}{\partial y^\rho} - y^\rho
\frac{\partial}{\partial y^\sigma} ) - n_\beta {\cal L} ^\rho_{\ \zeta}
\frac{\partial}{\partial n^\alpha} {\cal L}_\sigma^{\ \zeta}
(y^\sigma
\frac{\partial}{\partial y^\rho} - y^\rho
\frac{\partial}{\partial y^\sigma} )
\label{eqn:2.22}
\end{equation}
which was called $g(\lambda_{\alpha\beta})$ in \cite{II}.
So finally, we obtain
\begin{equation}
X_{\alpha\beta} =  {\cal L} _{\sigma\beta} {\cal L}^{\rho}_{\ \alpha} (y^\sigma
\frac{\partial}{\partial y^\rho} - y^\rho
\frac{\partial}{\partial y^\sigma} ) - n_\beta {\cal L} ^\rho_{\ \zeta}
\frac{\partial}{\partial n^\alpha} {\cal L}_\sigma^{\ \zeta}
(y^\sigma
\frac{\partial}{\partial y^\rho} - y^\rho
\frac{\partial}{\partial y^\sigma} )  + n_\beta
\frac{\partial}{\partial
n^\alpha} - n_\alpha \frac{\partial}{\partial n^\beta}
\label{eqn:2.23}
\end{equation}
which was called $ih_n (\lambda_{\alpha\beta})$ in \cite{II}.
It was shown that these generators satisfy the Lie algebra
of SL(2,C).  We will maintain the matrix notation for
${\cal M}_{\alpha\beta}$ so that (\ref{eqn:2.23}) may be written as
\begin{eqnarray}
X_{\alpha\beta} &=& [ {\cal L} (n) {\cal M}_{\alpha\beta} {\cal L} ^T ]^\mu_{\
\nu} y^\nu
\frac{\partial}{\partial y^\mu} -
[{\cal L} ({\cal M}_{\alpha\beta})^\rho_{\ \sigma} n^\sigma
\frac{\partial}{\partial n^\rho} {\cal L} ^T]^\mu_{\ \nu} y^\nu
\frac{\partial}{\partial y^\mu} -
({\cal M}_{\alpha\beta})^\rho_{\ \sigma} n^\sigma
\frac{\partial}{\partial n^\rho} \nonumber \\
&=& - y^T [ {\cal L} (n) {\cal M}_{\alpha\beta} {\cal L} ^T ]
\nabla_{\bf y} - y^T {\cal L} (n) [n^T {\cal M}_{\alpha\beta}
\nabla_{\bf n} ] {\cal L} ^T \nabla_{\bf y} - n^T {\cal M}_{\alpha\beta}
\nabla_{\bf n}
\label{eqn:2.24}
\end{eqnarray}
where $(\nabla_{\bf y})_\mu = \frac{\partial}{\partial y^\mu}$.
By defining the four matrices
\begin{equation}
S_\mu = {\cal L} \frac{\partial}{\partial n^\mu} {\cal L} ^T \qquad\qquad
\mu = 0,\cdots,3
\label{eqn:2.25}
\end{equation}
(which by (\ref{eqn:2.20}) are antisymmetric) equation
(\ref{eqn:2.24}) becomes
\begin{equation}
X_{\alpha\beta} = -\left\{ y^T [ {\cal L} (n) {\cal M}_{\alpha\beta} {\cal
L}^T ]
\nabla_{\bf y} + n_\mu ({\cal M}_{\alpha\beta})^{\mu\nu}
[ y^T S_\nu \nabla_{\bf y}
+ (\nabla_n)_\nu ] \right\}
\label{eqn:2.26}
\end{equation}

\section{Classical and Quantum Mechanics of the \newline
Generalized Phase Space}
\setcounter{equation}{0}

For classical dynamical systems whose potential depends only on
$\rho$ (given by (\ref{eqn:1.1})), we would like to write a
Lagrangian for the reduced ``one-body problem'' which includes an
explicit kinetic term for $n$.  A possible choice is
\begin{equation}
{\rm L} = \frac{1}{2} m \dot x^2 + \frac{1}{2} \lambda \dot n^2
- V(x^2)
\label{eqn:3.1}
\end{equation}
where $\lambda$ is a length scale required because $n$ is a unit vector.
Notice that when $\dot n =0$, the dynamics depend only on
$\dot x$ for fixed $n_\mu$ and so the relative coordinate
remains within RMS($n$).  Rewriting (\ref{eqn:2.9}) as,
\begin{equation}
\dot x = {\cal L}^T \dot y + \dot {\cal L} ^T y = {\cal L} ^T [ \dot y + {\cal
L} \dot {\cal L} ^T y]
\label{eqn:3.2}
\end{equation}
we may write (\ref{eqn:3.1}) in the form
\begin{equation}
{\rm L} = \frac{1}{2} m [ \dot y + {\cal L} \dot {\cal L} ^T y]^2 +
\frac{1}{2} \lambda \dot n^2 - V(x^2).
\label{eqn:3.3}
\end{equation}
By construction, (\ref{eqn:3.3}) is Lorentz invariant, and so
is invariant under the transformations induced by
(\ref{eqn:2.26}).  Therefore, applying Noether's theorem
\begin{eqnarray}
0=\delta {\rm L} &=& \frac{\partial{\rm L} }{\partial \xi ^i} \delta \xi^i
+ \frac{\partial{\rm L} }{\partial \dot \xi ^i} \delta \dot \xi^i
\nonumber \\
&=& \frac{\partial{\rm L} }{\partial \xi ^i} \delta \xi^i
+ \frac{\partial{\rm L} }{\partial \dot \xi ^i} \frac{d}{d \tau} \delta \xi^i
\nonumber \\
&=& \left [ \frac{\partial{\rm L} }{\partial \xi ^i}
- \frac{d}{d \tau}
\frac{\partial{\rm L} }{\partial \dot \xi ^i} \right]
\delta \xi^i + \frac{d}{d \tau} \left[ \frac{\partial{\rm L}}{\partial \dot \xi
^i}
\delta \xi^i \right],
\label{eqn:3.4}
\end{eqnarray}
where the first term vanishes for solutions to the
Euler-Lagrange equation,
and taking the variation to be
$\delta \xi^i = \frac{1}{2} \omega^{\alpha\beta}
X_{\alpha\beta} \ \xi^i$, one obtains the conservation law
\begin{equation}
\frac{d}{d \tau} [ {\rm p}^\mu X_{\alpha\beta} y_\mu +
\pi^\mu X_{\alpha\beta} n_\mu ] =0
\label{eqn:3.5}
\end{equation}
where
\begin{equation}
{\rm p}_\mu = \frac{\partial{\rm L} }{\partial \dot y^\mu} \qquad {\rm and}
\qquad \pi_\mu = \frac{\partial{\rm L} }{\partial \dot n^\mu} ,
\label{eqn:3.6}
\end{equation}
using the notation ${\rm p}_\mu$ for the variable conjugate to $y^\mu$
(for each $n^\mu$).  Since the variables $y^\mu$ are bounded by
the RMS parameterization (\ref{eqn:1.7}), the ${\rm p}_\mu$ are
symmetric but not self-adjoint.  These operators, however, occur in
combinations which have self-adjoint extensions.  We discuss
these questions elsewhere.
Using (\ref{eqn:2.26}) for $X_{\alpha\beta}$, (\ref{eqn:3.5}) becomes,
\begin{equation}
\frac{d}{d \tau} \{ y^T {\cal L} (n) {\cal M}_{\alpha\beta} {\cal L} ^T {\rm p}
+ n_\mu
({\cal M}_{\alpha\beta})^{\mu\nu} [ y^T S_\nu {\rm p} + \pi_\nu ] \}  =0.
\label{eqn:3.7}
\end{equation}
If we understand $\pi_\nu$, in the Poisson bracket sense,
as a derivative with respect to $n_\mu$, then
the quantum operators $h_n(\lambda_{\alpha\beta})$ of \cite{II}
now appear as classical constants of the motion
for the Lagrangian (\ref{eqn:3.1}).

To obtain the Hamiltonian, we first observe
that ${\cal L}$ depends on $\tau$ only through $n$, so
\begin{equation}
{\cal L} \dot {\cal L}^T = {\cal L} \left(\dot n^\nu \frac{\partial}{\partial
n^\nu}
{\cal L}^T\right) = \dot n^\nu S_\nu
\label{eqn:3.8}
\end{equation}
Applying (\ref{eqn:3.6}) to (\ref{eqn:3.3}),
\begin{equation}
{\rm p}_\mu = \frac{\partial{\rm L} }{\partial \dot y^\mu} = m[\dot
y_\mu + ({\cal L} \dot {\cal L}^T y)_\mu] \quad \Rightarrow
\quad {\rm p}=m[\dot y + \dot n^\nu S_\nu y]
\label{eqn:3.9}
\end{equation}
and
\begin{equation}
\pi_\mu = \frac{\partial{\rm L} }{\partial \dot n^\mu} = \lambda
\dot n_\mu + m[\dot y + \dot n^\nu S_\nu y]^T
\frac{\partial}{\partial \dot n^\mu}
[\dot y + \dot n^\nu S_\nu y] = \lambda \dot n_\mu - y^T S_\mu {\rm
p}
\label{eqn:3.10}
\end{equation}
where we used (\ref{eqn:3.9}) and the antisymmetry of $S_\mu$
to obtain (\ref{eqn:3.10}).
Equations (\ref{eqn:3.9}) and (\ref{eqn:3.10}) may be inverted
to eliminate $(\dot n , \dot y)$:
\begin{equation}
\dot n_\mu = \frac{1}{\lambda} [\pi_\mu + y^T S_\mu {\rm p}]
\label{eqn:3.11}
\end{equation}
and
\begin{equation}
\dot y = \frac{1}{m} {\rm p} - \dot n^\mu S_\mu y = \frac{1}{m} {\rm
p}-
\frac{1}{\lambda} [\pi^\mu + y^T S^\mu {\rm p}] S_\mu y
\label{eqn:3.12}
\end{equation}
which may be used to write the Hamiltonian as
\begin{eqnarray}
{\rm K} &=& \dot y \cdot {\rm p} + \dot n \cdot \pi - {\rm L} \nonumber \\
&=& {\rm p}^T (\frac{1}{m} {\rm p} - \frac{1}{\lambda} [\pi^\mu + y^T S^\mu
{\rm p}]
S_\mu y) + (\frac{1}{\lambda} [\pi_\mu + y^T S_\mu {\rm p}]) \pi^\mu -
\frac{1}{2} m (\frac{1}{m^2} {\rm p}^2) \nonumber \\
&   & \mbox{   } - \frac{1}{2} \lambda [ (\frac{1}{\lambda})^2
(\pi^\mu + y^T S^\mu {\rm p})(\pi_\mu + y^T S_\mu {\rm p})] + V \nonumber \\
&=& \frac{{\rm p}^2}{2m} + \frac{1}{2\lambda} (\pi^\mu +
y^T S^\mu {\rm p})(\pi_\mu + y^T S_\mu {\rm p}) + V
\label{eqn:3.13}
\end{eqnarray}

Since $S^\mu$ is antisymmetric, we may regard (\ref{eqn:3.13})
as a quantum Hamiltonian without ordering ambiguity in the
operator $y^T S^\mu {\rm p}$.  The Schr\"odinger equation is then
\begin{equation}
i\partial_\tau \psi = {\rm K} \psi = \left[ \frac{{\rm p}^2}{2m} +
\frac{1}{2\lambda} (\pi^\mu
+ y^T S^\mu {\rm p})(\pi_\mu + y^T S_\mu {\rm p}) + V \right] \psi ,
\label{eqn:3.14}
\end{equation}
where we take as quantum operators
\begin{equation}
{\rm p}_\mu = i \frac{\partial}{\partial y^\mu} \qquad \pi_\mu =
i \frac{\partial}{\partial n^\mu}
\label{eqn:3.15}
\end{equation}
We require that (\ref{eqn:3.14}) be locally gauge invariant
in the coordinate space $(n,y)$, that is, under transformations
of the form
\begin{equation}
\psi \longrightarrow e^{-ie\Theta(n,y)} \ \psi \ ;
\label{eqn:3.16}
\end{equation}
this can be accomplished through the minimal coupling prescription
\begin{equation}
{\rm p}_\mu \longrightarrow {\rm p}_\mu -  e{\rm A}^{(n)}_\mu \qquad
\pi_\mu \longrightarrow \pi_\mu -  e\chi_\mu
\label{eqn:3.17}
\end{equation}
together with the requirement that under gauge transformation
\begin{equation}
{\rm A}^{(n)}_\mu \longrightarrow {\rm A}^{(n)}_\mu +
\frac{\partial}{\partial y^\mu}
\Theta \qquad
\chi_\mu \longrightarrow \chi_\mu + (\frac{\partial}{\partial n^\mu}
+ y^T S_\mu \nabla_{\bf y}) \Theta.
\label{eqn:3.18}
\end{equation}
Note that ${\rm A}^{(n)}_\mu$ transforms under O(3,1) as an
induced (over O(2,1))
representation; it transforms as ${\rm p}_\mu$ under Lorentz
transformations (i.e., under the O(2,1) little group) and so,
since the Maxwell equations are Lorentz invariant,
it satisfies the Maxwell equation in the $y^\mu$
variables.  Under gauge transformation,
\begin{equation}
({\rm p} -e{\rm A}^{(n)\prime})e^{-ie\Theta}\psi =
e^{-ie\Theta}({\rm p}+e\nabla_{\bf y}
\Theta - e{\rm A}^{(n)\prime})\psi = e^{-ie\Theta}
({\rm p} - e{\rm A}^{(n)} )\psi
\label{eqn:3.19}
\end{equation}
and
\begin{eqnarray}
(\pi_\mu + y^T S_\mu {\rm p} -e\chi'_\mu )e^{-ie\Theta}\psi &=&
e^{-ie\Theta}
(\pi_\mu + y^T S_\mu {\rm p} + e \frac{\partial}{\partial n^\mu} \Theta
+ e y^T S_\mu \nabla_{\bf n} \Theta - e\chi'_\mu )\psi \nonumber \\
&=& e^{-ie\Theta} (\pi_\mu + y^T S_\mu {\rm p} -e\chi_\mu )\psi ,
\label{eqn:3.20}
\end{eqnarray}
so that the gauge invariant form of (\ref{eqn:3.14}) is
\begin{equation}
i\partial_\tau \psi = {\rm K} \psi = \left[ \frac{1}{2m}
({\rm p}-e{\rm A}^{(n)})^2+
\frac{1}{2\lambda} (\pi^\mu + y^T S^\mu {\rm p} -e\chi^\mu )(\pi_\mu +
y^T S_\mu {\rm p} -e\chi_\mu ) + V \right] \psi
\label{eqn:3.21}
\end{equation}

Consider the derivative operator which acts on $\Theta(n,y)$
in the transformation of the gauge field $\chi_\mu$ in
(\ref{eqn:3.18}).  We denote this operator by
\begin{equation}
D_\mu =  (\nabla_n)_\mu + y^T S_\mu \nabla_{\bf y}
\label{eqn:3.22}
\end{equation}
and we notice that $D_\mu$ also appears in the Lorentz
generators $X_{\alpha\beta}$ (\ref{eqn:2.26}).  From
(\ref{eqn:3.11}) we see that $D_\mu$ may be regarded
as the quantum operator corresponding to $\lambda \dot n$.
Using (\ref{eqn:3.22})
in (\ref{eqn:2.26}), the generators assume the simpler form
\begin{eqnarray}
X_{\alpha\beta} &=& -\{ y^T [ {\cal L} (n)
{\cal M}_{\alpha\beta} {\cal L} ^T ] \nabla_{\bf y}
+ n_\mu ({\cal M}_{\alpha\beta})^{\mu\nu} D_\nu \}
\nonumber \\
&=&  -\{ x^T {\cal M}_{\alpha\beta} \nabla_{\bf x} + n_\mu
({\cal M}_{\alpha\beta})^{\mu\nu} D_\nu \}
\label{eqn:3.23}
\end{eqnarray}
which, in light of (\ref{eqn:3.11}) and the definitions of
${\rm p}_\mu$ and $\pi_\mu$, suggests the analog
\begin{equation}
X_{\alpha\beta} \sim i \ [x^T {\cal M}_{\alpha\beta} (m\dot x) +
\ n^T {\cal M}_{\alpha\beta} (\lambda \dot n)].
\label{eqn:3.24}
\end{equation}
In fact, using (\ref{eqn:3.9}) and (\ref{eqn:3.11}) in (\ref{eqn:3.7}),
we find for the classical conservation law, that
\begin{eqnarray}
\frac{d}{d \tau} \left\{ y^T {\cal L} (n) {\cal M}_{\alpha\beta} {\cal L} ^T p
\right.
&+& \left. n_\mu
({\cal M}_{\alpha\beta})^{\mu\nu} [ y^T S_\nu p + \pi_\nu ] \right\}  =0
\nonumber \\
&=& \frac{d}{d \tau} \left\{ m \ y^T {\cal L} (n) {\cal M}_{\alpha\beta} {\cal
L}^T
[\dot y + \dot n^\nu S_\nu y]
+ n^T ({\cal M}_{\alpha\beta}) [ \lambda \dot n] \right\}
\nonumber \\
&=& \frac{d}{d \tau} \left\{ m \ y^T {\cal L} (n) {\cal M}_{\alpha\beta} {\cal
L}^T
[\dot y + {\cal L}\dot{\cal L}^T y]
+ n^T ({\cal M}_{\alpha\beta}) [ \lambda \dot n] \right\}
\nonumber \\
&=& \frac{d}{d \tau} \left\{ m \ x^T {\cal M}_{\alpha\beta}
[{\cal L}^T \dot y + \dot{\cal L}^T y]
+ n^T ({\cal M}_{\alpha\beta}) [ \lambda \dot n] \right\}
\nonumber \\
&=& \frac{d}{d \tau} \left\{ x^T {\cal M}_{\alpha\beta} [m \dot x]
+ n^T ({\cal M}_{\alpha\beta}) [ \lambda \dot n] \right\}
\label{eqn:3.25}
\end{eqnarray}
providing the generators with the form of a generalized
angular momentum in terms of the relative Minkowski variables
and the frame orientation variables.

The Hamiltonian (\ref{eqn:3.13}) also assumes a simple form when
expressed in terms of (\ref{eqn:3.22}):
\begin{equation}
{\rm K} = - \frac{1}{2m}\ \frac{\partial}{\partial y^\mu}
\frac{\partial}{\partial y_\mu} - \frac{1}{2\lambda} D_\mu D^\mu  + V.
\label{eqn:3.26}
\end{equation}

Suppose that a function $f(n,y)$ is defined in such a way that
its dependence on $n$ is only through ${\cal L}(n)^T y$ (which is to
say that $f$ is a function of $x$ alone, even as $n$ varies in
$\tau$).  Then we find that
\begin{equation}
\frac{\partial}{\partial y^\mu} f = \left.\frac{df}{d\xi^\alpha}
\right|_{\xi={\cal L}(n)^T y} \frac{\partial}{\partial y^\mu} ({\cal
L}_\beta^{\ \alpha} y^\beta) =
{\cal L}_\mu^{\ \alpha} \left.\frac{df}{d\xi^\alpha}
\right|_{\xi={\cal L}(n)^T y}
\label{eqn:3.27}
\end{equation}
and
\begin{equation}
\frac{\partial}{\partial n^\mu} f = \left.\frac{df}{d\xi^\alpha}
\right|_{\xi={\cal L}(n)^T y} \frac{\partial}{\partial n^\mu} ({\cal
L}_\beta^{\ \alpha} y^\beta)
\label{eqn:3.28}
\end{equation}
so that
\begin{eqnarray}
D_\mu f &=& \left(\frac{\partial}{\partial n^\mu} + y^T S_\mu
\nabla_{\bf y} \right) f
\nonumber \\
&=& \left[\frac{\partial}{\partial n^\mu} + y_\beta {\cal L}^\beta
_{\ \gamma} (\frac{\partial}{\partial n^\mu}
{\cal L} ^{\alpha\gamma}) \frac{\partial}{\partial
y^\alpha}\right] f \nonumber \\
&=&  \left.\frac{df}{d\xi^\sigma} \right|_{\xi={\cal L}(n)^T y}
y^\beta
\left[
\frac{\partial}{\partial n^\mu}
{\cal L}_\beta^{\ \sigma}
+ {\cal L}_\beta^{\ \gamma}
( \frac{\partial}{\partial n^\mu} {\cal L} ^{\alpha}_{\ \gamma})
{\cal L}_\alpha^{\ \sigma}
\right] \nonumber \\
&=&  \left.\frac{df}{d\xi^\sigma} \right|_{\xi={\cal L}(n)^T y}
y^\beta
\left[
\frac{\partial}{\partial n^\mu}
{\cal L}_\beta^{\ \sigma} + {\cal L}_\beta^{\ \gamma}
({\cal L}^T)^\sigma_{\ \alpha} \frac{\partial}{\partial n^\mu}
{\cal L}^{\alpha}_{\ \gamma}
\right] \nonumber \\
&=& \left.\frac{df}{d\xi^\sigma} \right|_{\xi={\cal L}(n)^T y} y^\beta
\left[ \frac{\partial}{\partial n^\mu} {\cal L}_\beta^{\ \sigma}
- {\cal L}_\beta^{\ \gamma} {\cal L} ^{\alpha}_{\ \gamma}
\frac{\partial}{\partial n^\mu} {\cal L}_\alpha^{\ \sigma} \right]
\nonumber \\ &\equiv& 0
\label{eqn:3.29}
\end{eqnarray}
where we have used (\ref{eqn:2.20}).
Thus, $D_\mu$ acts as a kind of covariant derivative which
vanishes on functions of $x$ alone.
In particular, $D_\mu$ vanishes on the eigenstates discussed in
\cite{I} and \cite{II}, in which case the Hamiltonian
(\ref{eqn:3.13}, \ref{eqn:3.26}) reduces to the RMS Hamiltonian
discussed in \cite{I}.  The dynamical effects that we shall
discuss in the next section are associated with the evolution of
the wave function of the system to a form which does not depend
only on $x^\mu$.

Notice also that
\begin{eqnarray}
\dot n^\mu D_\mu &=& \left(\dot n^\mu \frac{\partial}{\partial n^\mu}
+ y^T \dot n^\mu S_\mu \nabla_{\bf y} \right)
\nonumber \\
&=& \left(\dot n \cdot \nabla_{\bf n}  - y^T \dot {\cal L} {\cal L}^T
\nabla_{\bf y}  \right)
\nonumber \\
&=& \left(\dot n \cdot \nabla_{\bf n}  -
\left[ \frac{d}{d\tau} ({\cal L}^T y) - \dot y^T {\cal L} \right]
{\cal L}^T \nabla_{\bf y}  \right)
\nonumber \\
&=& \left(\dot n \cdot \nabla_{\bf n}  +
\dot y \cdot \nabla_{\bf y}
- \dot x \cdot \nabla_{\bf x}  \right)
\label{eqn:3.30}
\end{eqnarray}
We may rewrite this expression as
\begin{equation}
dx \cdot \nabla_{\bf x} + dn^\mu D_\mu = dy \cdot
\nabla_{\bf y} + dn \cdot \nabla_{\bf n}
\label{eqn:3.31}
\end{equation}
which shows in yet another way that $\nabla_{\bf x}$ and $D_\mu$
generate the changes induced by $dx$ and $dn$ (with $x^\mu$ held
constant), just
as $\nabla_{\bf y}$ and $\nabla_{\bf n}$ generate the
changes induced by $dy$ and $dn$ (with $y^\mu$ held
constant).

It will be useful to examine the classical Lagrangian in the
presence of the fields ${\rm A}^{(n)}_\mu$ and $\chi_\mu$, which we may find
by treating the Hamiltonian in (\ref{eqn:3.21}) as a classical
functional and evaluating
\begin{equation}
\dot n^\mu = \frac{\partial}{\partial \pi_\mu} {\rm K} =
\frac{1}{\lambda } (\pi^\mu + y^T S^\mu {\rm p} -e \chi^\mu)
\label{eqn:3.32}
\end{equation}
and
\begin{eqnarray}
\dot y_\mu &=& \frac{\partial}{\partial {\rm p}^\mu} {\rm K} =
\frac{1}{m} ({\rm p}_\mu -e {\rm A}^{(n)}_\mu) + \frac{1}{\lambda }
(\pi_\nu + y^T S_\nu {\rm p} -e \chi_\nu) \frac{\partial}{\partial \pi^\mu}
(y^T S^\nu {\rm p})
\nonumber \\
&=& \frac{1}{m} ({\rm p}_\mu -e {\rm A}^{(n)}_\mu) - \dot n_\nu
(S^\nu)_{\mu\sigma} y^\sigma .
\label{eqn:3.33}
\end{eqnarray}
Recalling (\ref{eqn:3.8}), we find that
\begin{eqnarray}
{\rm L} &=&{\rm p} \cdot \dot y + \pi \cdot \dot n - {\rm K}
\nonumber \\
&=& \frac{1}{2} m [ \dot y + {\cal L} \dot {\cal L} ^T y]^2 +
\frac{1}{2} \lambda \dot n^2
+ e [ (\dot y + {\cal L} \dot {\cal L} ^T y) \cdot {\rm A}^{(n)} + \dot n \cdot
\chi ]
- V(x^2).
\label{eqn:3.34}
\end{eqnarray}
 From (\ref{eqn:3.2}), we have
\begin{equation}
\dot y + {\cal L} \dot {\cal L} ^T y = {\cal L} \dot x,
\label{eqn:3.35}
\end{equation}
so that we may write (\ref{eqn:3.34}) in the form
\begin{equation}
{\rm L} = \frac{1}{2} m \dot x^2 + \frac{1}{2} \lambda \dot n^2
+ e [ \dot x \cdot ({\cal L}^T {\rm A}^{(n)}) + \dot n \cdot \chi ] - V(x^2).
\label{eqn:3.36}
\end{equation}
In order for L to be a Lorentz scalar, ${\cal L}^T {\rm A}^{(n)}$ must
transform under
the full Lorentz group O(3,1).  Since ${\rm A}^{(n)}$ was introduced as a
field which transforms under the O(2,1) little group, we may write
\begin{equation}
{\rm A}^{(n)\prime} = D^{-1} (\Lambda , n) {\rm A}^{(n)} =
{\cal L} (\Lambda n) \; \Lambda \; {\cal L}^T (n) {\rm A}^{(n)}
\Longrightarrow {\cal L}^T \; (\Lambda n) {\rm A}^{(n)\prime} =
\Lambda \; {\cal L}^T (n) {\rm A}^{(n)}
\label{eqn:3.37}
\end{equation}
verifying that the combination ${\cal L}^T {\rm A}^{(n)}$ transforms as a four
vector under $\Lambda$.

\section{The Zeeman Effect}
\setcounter{equation}{0}

In \cite{II}, the spacelike vector $n$ played no particular
role in the dynamics and could be chosen arbitrarily, because
the systems under discussion were O(3,1)-symmetric and no
direction in spacetime was intrinsic to the problem (other than
the axis of the bound state).  That situation generalizes the
nonrelativistic spherically symmetric central force problem, in
which the absence of a preferred direction in space leads to
the degeneracy of the energy spectrum with respect to the magnetic
quantum number (which characterizes the orientation of the
angular momentum).  In \cite{selrul}, it was shown that the
vector $n$ plays a role in dipole radiation from the bound
state, because conservation of angular momentum and the spin-1
nature of the electromagnetic field impose an orientation
dependence on the interaction.  Thus, the photon carries off
spin provided by the bound state transition, and that
transition depends on the orientation of the angular momentum
of the state (determined by $n$) and the photon polarization.

In the Zeeman effect, one lifts the degeneracy of the bound
state spectrum by placing the state in a constant external
magnetic field, which interacts with the magnetic
moment (angular momentum) of the system and thereby provides
a preferred direction in space.  In
the semiclassical picture, the atom will tend to rotate.  The
interaction angular momentum is intimately connected with the
rotation generators, and for the bound states discussed here,
these generators are elements of the rotation subgroup of the
induced representation of O(3,1).  Since the rotation group
O(3) $\subset$ O(3,1) acts on the vector $n$ as well as the RMS
variables $y$, the relativistic Zeeman effect can clearly
only be described in the context of a theory which explicitly
permits the generators to act
directly on all the variables in the theory.  In this section,
we provide such a description in the context of the Hamiltonian
theory given in the Section 4.

In the nonrelativistic case, the Zeeman effect is obtained
as a first order perturbation of the hydrogen atom bound state,
by a vector potential
\begin{equation}
{\bf A}({\bf r}) = -\frac{1}{2} {\bf B} \times {\bf r}
\label{eqn:4.1}
\end{equation}
which leads to the constant magnetic field
\begin{equation}
(\nabla \times {\bf A} )^i = \epsilon^{ijk}
\frac{\partial}{\partial r^j}
(-\frac{1}{2} \epsilon_{klm} B_l r_m) = B^i.
\label{eqn:4.2}
\end{equation}
The Hamiltonian becomes
\begin{eqnarray}
{\rm H} &=&  \frac{1}{2m} ({\bf p}-e{\bf A})^2 + V  \nonumber \\
&=&  \frac{{\bf p}^2}{2m}  + V + \frac{e}{2m}
({\bf p} \cdot {\bf A} + {\bf A} \cdot {\bf p}) +o(e^2) \nonumber \\
&=& {\rm H}_0 + \frac{e}{m} {\bf A} \cdot {\bf p} +o(e^2) \nonumber \\
&=& {\rm H}_0 - \frac{e}{2m} ({\bf B} \times {\bf r} ) \cdot {\bf p}
 +o(e^2) \nonumber \\
&=& {\rm H}_0 - \frac{e}{2m} {\bf B} \cdot ({\bf r} \times {\bf p})
 +o(e^2) \nonumber \\
&=& {\rm H}_0 - \frac{e}{2m} {\bf B} \cdot {\bf L} +o(e^2)
\label{eqn:4.3}
\end{eqnarray}
where ${\bf L} = {\bf r} \times {\bf p}$ is the angular
momentum operator.  Thus taking ${\bf B}$ in the direction of
the diagonal angular momentum operator (usually the $z$-axis),
the observed Zeeman splitting is obtained from (\ref{eqn:4.3})
as
\begin{equation}
E_{ln} \longrightarrow E_{lnq} = E_{ln} - \frac{eB}{2m} q.
\label{eqn:4.4}
\end{equation}
where $q$ is the eigenvalue of the operator $L_z$.

In Section 3, we introduced two gauge compensation
fields, ${\rm A}^{(n)}_\mu$ and $\chi_\mu$, required to make the
Hamiltonian (\ref{eqn:3.13}) locally gauge invariant.  However,
we now argue that just as $n$ and $y$ transform under
inequivalent representations of the Lorentz group ($y$ transforms
under the O(2,1) little group induced by the action of the full
O(3,1)), so ${\rm A}^{(n)}_\mu$
and $\chi_\mu$ must be seen as inequivalent representations of
the usual U(1) gauge group of electromagnetism.  In the full
spacelike region, a constant electromagnetic
field, $F^{\mu\nu}$, can be represented through the vector potential
\begin{equation}
A^\mu (x) = -\frac{1}{2} F^{\mu\nu} x_\nu.
\label{eqn:4.5}
\end{equation}
We now restrict the support of $A^\mu$ to $x\in$ RMS($n$)
and express the vector potential as a vector oriented with
RMS(${\mathaccent'27n}$) by writing
\begin{equation}
{\rm A}^{(n)}_\mu (y)= {\cal L}_{\mu\nu} A^\nu ({\cal L}^T y)
=-\frac{1}{2} {\cal L}_{\mu\nu} F^\nu_{\ \sigma}
{\cal L}_\lambda^{\ \sigma} y^\lambda
=-\frac{1}{2} ({\cal L} F {\cal L}^T y)_\mu.
\label{eqn:4.6}
\end{equation}
For the field $\chi_\mu$, we choose (note
that $n$ undergoes Lorentz transform in the same way as $x$),
\begin{equation}
\chi_\mu (n) = b^2 \ A_\mu (n)
= -\frac{b^2}{2} \ F^\nu_{\ \sigma} \ n^\sigma
\label{eqn:4.7}
\end{equation}
(here $b$ is another length scale, required since $A_\mu (x)$ has
units of length$^{-1}$, so $F^\nu_\sigma$ must have units of
length$^{-2}$, but $\chi_\mu$ must be without units) and we use
(\ref{eqn:4.6}) and (\ref{eqn:4.7}) in the Schr\"odinger
equation (\ref{eqn:3.21}).
\begin{eqnarray}
i\partial_\tau \psi &=&  \left[ \frac{1}{2m} ({\rm p}-e{\rm A}^{(n)})^2+
\frac{1}{2\lambda} (\pi^\mu + y^T S^\mu {\rm p} -e\chi^\mu )
(\pi_\mu + y^T S_\mu {\rm p} -e\chi_\mu ) + V \right] \psi \nonumber \\
&=& \left[
\frac{1}{2m} {\rm p}^2 - \frac{e}{2m} ({\rm p} \cdot {\rm A}^{(n)} + {\rm
A}^{(n)} \cdot {\rm p})
+ \frac{1}{2\lambda} (\pi^\mu + y^T S^\mu {\rm p} )^2 - \right.
\nonumber \\
& & \mbox{\quad }\left. \frac{e}{2\lambda} [(\pi^\mu + y^T S^\mu {\rm
p}
)\chi_\mu +
\chi^\mu (\pi_\mu + y^T S_\mu {\rm p}) ]
+ V +o(e^2) \right] \psi \nonumber \\
&=&  \left[ \frac{1}{2m} {\rm p}^2
+ \frac{1}{2\lambda} (\pi^\mu + y^T S^\mu {\rm p} )^2
+ V \right.  \nonumber \\
& & \mbox{\qquad }\left. - e [\frac{1}{m} {\rm A}^{(n)} \cdot {\rm
p}+
\frac{1}{\lambda} \chi^\mu (\pi_\mu + y^T S_\mu {\rm p}) ]
+ o(e^2) \right] \psi
\label{eqn:4.8}
\end{eqnarray}
where the first three terms of (\ref{eqn:4.8}) are the
unperturbed Hamiltonian ${\rm K}_0$.

The perturbation term to order $o(e)$, is
\begin{eqnarray}
-e [\frac{1}{m} {\rm A}^{(n)} \cdot {\rm p} &+& \frac{1}{\lambda} \chi^\mu
(\pi_\mu + y^T S_\mu {\rm p}) ]
\nonumber \\
&=& - e [\frac{1}{m} {\rm A}^{(n)T}  {\rm p} +
\frac{1}{\lambda} [\chi^T \pi + y^T (S\cdot\chi){\rm p}]
\nonumber \\
&=& -\frac{e}{2} [\frac{1}{m} ({\cal L} F {\cal L}^T y)^T {\rm p}
+ \frac{b^2}{\lambda} F^\mu_{\ \nu} n^\nu (\pi_\mu + y^T S_\mu {\rm p})]
\nonumber \\
&=& \frac{e}{2m} [y^T {\cal L} F {\cal L}^T {\rm p}
+ \frac{mb^2}{\lambda} n_\nu F^{\nu\mu} (\pi_\mu + y^T S_\mu {\rm p})].
\label{eqn:4.9}
\end{eqnarray}
We now expand the electromagnetic field tensor on the basis of
four by four antisymmetric tensors given by the Lorentz
generators ${\cal M}^{\mu\nu}$.  Thus,
\begin{equation}
F = \frac{1}{2} F_{\mu\nu} {\cal M}^{\mu\nu}
\label{eqn:4.10}
\end{equation}
may be verified through
\begin{equation}
(F)^{\alpha\beta} = \frac{1}{2} F_{\mu\nu} ({\cal M}^{\mu\nu})^{\alpha\beta}
= \frac{1}{2} F_{\mu\nu} (g^{\mu\alpha}g^{\nu\beta} -
g^{\mu\beta}g^{\nu\alpha}) = F^{\alpha\beta}.
\label{eqn:4.11}
\end{equation}
Using (\ref{eqn:4.10}) in (\ref{eqn:4.9}) we find that
the perturbation term to order $o(e)$ becomes
\begin{equation}
\frac{e}{4m} F_{\alpha\beta} [y^T {\cal L} {\cal M}^{\alpha\beta}
{\cal L}^T {\rm p}
+ \frac{mb^2}{\lambda} n_\mu ({\cal M}^{\alpha\beta})^{\mu\nu}
(\pi_\nu + y^T S_\nu {\rm p})]
\label{eqn:4.12}
\end{equation}
We note that if $\lambda / b^2 = m$, then we may write the first
order perturbation (using (\ref{eqn:3.23})) as
\begin{equation}
\frac{e}{4m} F_{\alpha\beta} [y^T {\cal L}
{\cal M}^{\alpha\beta} {\cal L}^T {\rm p}
+ n_\mu ({\cal M}^{\alpha\beta})^{\mu\nu}
(\pi_\nu + y^T S_\nu {\rm p})] = \frac{e}{4m} F_{\alpha\beta}
X^{\alpha\beta} .
\label{eqn:4.13}
\end{equation}

For $F^{\mu\nu}F_{\mu\nu} = 2({\bf B}^2 - {\bf E}^2) >0$, there
exists a frame for which the interaction is purely magnetic.  In
such a frame, the perturbation becomes
\begin{equation}
\frac{e}{4m} F_{\alpha\beta} X^{\alpha\beta} =
\frac{e}{4m} F_{ij} X^{ij} = \frac{e}{4m} \epsilon_{ijk} B^k
X^{ij} = \frac{e}{2m} B^k \left[\frac{1}{2} \epsilon_{ijk}
X^{ij}\right]
= \frac{e}{2m} B^k h(\lambda_k)
\label{eqn:4.14}
\end{equation}

where $h(\lambda_k)$ are
the three conserved generators of the SU(2) rotation subgroup of
SL(2,C) for the phase space $\{(n,y);(\pi,{\rm p})\}$, that is, the
angular momentum operator for the eigenstates of the induced
representation.  Notice that in the matrix element for unperturbed
eigenstates, the second terms of (\ref{eqn:4.9}) vanishes, so the
relativistic Zeeman effect does not depend upon the values of
$\lambda$ or $b$.

In \cite{II}, the diagonal angular momentum operator is
$L_1(n) = h(\lambda_1)=-i\partial/\partial \gamma$, and
so if we take ${\bf B} =
B(1,0,0)$ then we find that
\begin{equation}
{\rm K}_0 \quad \longrightarrow \quad {\rm K} = {\rm K}_0 -
\frac{eB}{2m} h(\lambda_1)
\label{eqn:4.15}
\end{equation}
splits the mass levels of the bound states according to
\begin{equation}
E_{\ell n} \quad \longrightarrow \quad E_{\ell n} - \frac{eB}{2m} q
\label{eqn:4.16}
\end{equation}
In going from (\ref{eqn:4.15}) to (\ref{eqn:4.16}), we have
used the fact that the unperturbed Hamiltonian of
(\ref{eqn:4.8}) reduces to the the unperturbed Hamiltonian of
\cite{II}.
Equation (\ref{eqn:4.16}) further justifies the conclusion reached
in \cite{selrul} that $q$ is the magnetic quantum number.
Moreover, the manifest covariance of the formalism guarantees
that the splitting of the spectrum will be independent of the
observer.  We observe that if $F^{\mu\nu}F_{\mu\nu} <0$, we may
find a frame in which the interaction is purely electric, leading
to a covariant formulation of the Stark effect.  Since the
electric field couples to the boost generators (which reduce to
the position operator in the nonrelativistic limit) and these
generators are not diagonal in this representation, the Stark
effect remains formally (one really has only a resonance spectrum;
the bound states are destroyed by the non-compact generator)
a second order perturbation, and we will discuss
it elsewhere.

%
%%%%%%%%%%%%%%%%%%%%%%%%%% REFERENCES %%%%%%%%%%%%%%%%%%%%%%%%%%%%%
%

\end{document}